\documentclass[12pt,prd,aps,amssymb,amsmath,tightenlines,showpacs]{revtex4}
\usepackage{psfig}

\usepackage{graphics}
\usepackage[pdftex]{graphicx}

\begin{document}


\title{MUST A HAMILTONIAN BE HERMITIAN?}

\author{Carl~M.~Bender$^{*}$, Dorje~C.~Brody$^{\dagger}$, and
Hugh F. Jones$^{\dagger}$}

\affiliation{${}^{*}$Department of Physics, Washington University,
St. Louis MO 63130, USA}
\affiliation{${}^{\dagger}$Blackett
Laboratory, Imperial College, London SW7 2BZ, UK}

\date{\today}

\begin{abstract}
A consistent physical theory of quantum mechanics can be built on
a complex Hamiltonian that is not Hermitian but instead satisfies
the physical condition of space-time reflection symmetry (${\cal
PT}$ symmetry). Thus, there are infinitely many new Hamiltonians
that one can construct that might explain experimental data. One
would think that a quantum theory based on a non-Hermitian
Hamiltonian violates unitarity. However, if ${\cal PT}$ symmetry
is not broken, it is possible to use a previously unnoticed
physical symmetry of the Hamiltonian to construct an inner product
whose associated norm is positive definite. This construction is
general and works for any ${\cal PT}$-symmetric Hamiltonian. The
dynamics is governed by unitary time evolution. This formulation
does not conflict with the requirements of conventional quantum
mechanics. There are many possible observable and experimental
consequences of extending quantum mechanics into the complex
domain, both in particle physics and in solid state physics.
\end{abstract}


\maketitle

\section{Introduction}
\label{s1} In this paper we present an alternative to the one of
the standard axioms of quantum mechanics; namely, that the
Hamiltonian $H$, which incorporates the symmetries and specifies
the dynamics of a quantum theory, must be Hermitian:
$H=H^\dagger$. It is commonly believed that the Hamiltonian must
be Hermitian in order to ensure that the energy spectrum (the
eigenvalues of the Hamiltonian) is real and that the time
evolution of the theory is unitary (probability is conserved in
time). Although this axiom is sufficient to guarantee these
desired properties, we argue here that it is not necessary. We
believe that the condition of Hermiticity is a mathematical
requirement whose physical basis is somewhat remote and obscure.
We demonstrate here that there is a simpler and more physical
alternative axiom, which we refer to as space-time reflection
symmetry (${\cal PT}$ symmetry): $H= H^{\cal PT}$. This symmetry
allows for the possibility of non-Hermitian and complex
Hamiltonians but still leads to a consistent theory of quantum
mechanics.

We also show that because ${\cal PT}$ symmetry is an alternative
condition to Hermiticity it is now possible to construct
infinitely many new Hamiltonians that would have been rejected in
the past because they are not Hermitian. An example of such a
Hamiltonian is $H=p^2+ix^3$. It should be emphasized that we do
not regard the condition of Hermiticity as wrong. Rather, the
condition of ${\cal PT}$ symmetry offers the possibility of
studying new and interesting quantum theories.

Let us recall the properties of the space reflection (parity)
operator ${\cal P}$ and the time-reflection operator ${\cal T}$.
The parity operator ${\cal P}$ is {\it linear} and has the effect
$$ p\to-p \quad {\rm and} \quad x\to-x.$$
The time-reversal operator ${\cal T}$ is antilinear and has the
effect
$$p\to-p, \quad x\to x, \quad {\rm and} \quad i\to-i.$$
Note that ${\cal T}$ changes the sign of $i$ because, like the
parity operator, it preserves the fundamental commutation relation
of quantum mechanics, $[x,p] =i$, known as the Heisenberg algebra
\footnote{The Heisenberg-Weyl algebra is a real three-dimensional
Lie algebra whose generators satisfy the commutation relations
$[e_1,e_2]=e_3$, $[e_1,e_3]=[e_2,e_3]=0$. To recover the
Heisenberg commutation relations we set $e_1=i(\hbar)^{-1/2}p$,
$e_2=i(\hbar)^{-1/2}x$, and $e_3=i$.}.

It is easy to construct infinitely many Hamiltonians that are not
Hermitian but do possess ${\cal PT}$ symmetry. For example,
consider the one-parameter family of Hamiltonians
\begin{equation}
H=p^2+x^2(ix)^\epsilon\qquad (\epsilon~{\rm real}). \label{e1}
\end{equation}
Note that while $H$ in (\ref{e1}) is not symmetric under ${\cal
P}$ or ${\cal T}$ separately, it is invariant under their combined
operation. We say that such Hamiltonians possess space-time
reflection symmetry. Other examples of complex Hamiltonians having
${\cal PT}$ symmetry are $H=p^2+x^4(ix)^\epsilon$,
$H=p^2+x^6(ix)^\epsilon$, and so on \cite{r1}\footnote{These
classes of Hamiltonians are all {\it different}. For example, the
Hamiltonian obtained by continuing $H$ in (\ref{e1}) along the
path $\epsilon:\,0\to8$ has a different spectrum from the
Hamiltonian that is obtained by continuing $H=p^2+x^6(ix)^
\epsilon$ along the path $\epsilon:\,0\to4$. This is because the
boundary conditions on the eigenfunctions are different.}.

The class of ${\cal PT}$-symmetric Hamiltonians is larger than and
includes real symmetric Hermitians because any real symmetric
Hamiltonian is automatically ${\cal PT}$-symmetric. For example,
consider the real symmetric Hamiltonian $H=p^2+x^2+2x$. This
Hamiltonian is time-reversal symmetric, but according to the usual
definition of space reflection for which $x\to-x$, this
Hamiltonian appears not to have ${\cal PT}$ symmetry. However,
recall that the parity operator is defined only up to unitary
equivalence \cite{BBHM}. In this example, if we express the
Hamiltonian in the form $H=p^2+(x+1)^2-1$, then it is evident that
$H$ is ${\cal PT}$ symmetric, provided that the parity operator
performs a space reflection about the point $x=-1$ rather than
$x=0$. See Ref. \cite{BBB} for the general construction of the
relevant parity operator.

\begin{figure}[th]
{\centerline{\psfig{file=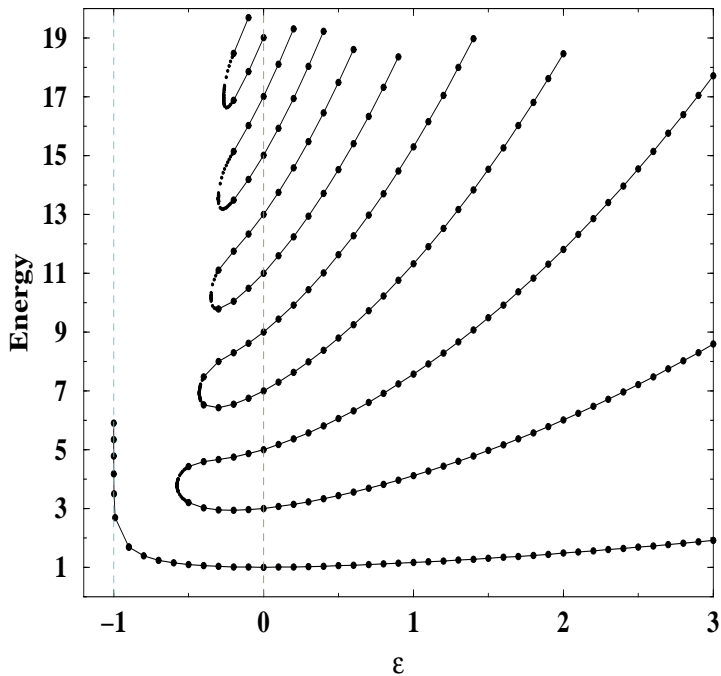,width=8cm,angle=0}}}
\caption{Energy levels of the Hamiltonian $H=p^2+x^2(ix)^\epsilon$
as a function of the parameter $\epsilon$. There are three
regions: When $\epsilon\geq0$, the spectrum is real and positive
and the energy levels rise with increasing $\epsilon$. The lower
bound of this region, $\epsilon=0$, corresponds to the harmonic
oscillator, whose energy levels are $E_n=2n+1$. When
$-1<\epsilon<0$, there are a finite number of real positive
eigenvalues and an infinite number of complex conjugate pairs of
eigenvalues. As $\epsilon$ decreases from $0$ to $-1$, the number
of real eigenvalues decreases; when $\epsilon\leq-0.57793$, the
only real eigenvalue is the ground-state energy. As $\epsilon$
approaches $-1^+$, the ground-state energy diverges. For $\epsilon
\leq-1$ there are no real eigenvalues.} \label{f1}
\end{figure}

Five years ago it was discovered that with properly defined
boundary conditions the spectrum of the Hamiltonian $H$ in
(\ref{e1}) is {\em real and positive} when $\epsilon\geq0$
\cite{r2}. The spectrum is partly real and partly complex when
$\epsilon<0$. The eigenvalues have been computed numerically to
very high precision, and the real eigenvalues are plotted as
functions of $\epsilon$ in Fig.~\ref{f1}.

We say that the ${\cal PT}$ symmetry of a Hamiltonian $H$ is {\it
unbroken} if all of the eigenfunctions of $H$ are simultaneously
eigenfunctions of ${\cal PT}$ \footnote{In general, if a system is
defined by an equation that possesses a discrete symmetry, the
solution to this equation need not exhibit that symmetry. For
example, the differential equation ${\ddot y} (t)=y(t)$ is
symmetric under the discrete time-reversal symmetry $t\to-t$. Note
that the solutions $y(t)=e^t$ and $y(t)=e^{-t}$ do not exhibit
this time-reversal symmetry while the solution $y(t)=\cosh(t)$ is
time-reversal symmetric. The same is true with a system whose
Hamiltonian is ${\cal PT}$ symmetric. Even if the Schr\"odinger
equation and the corresponding boundary conditions are ${\cal PT}$
symmetric, the wave function that solves the Schr\"odinger
equation boundary value problem may not be symmetric under
space-time reflection. When the solution exhibits ${\cal PT}$
symmetry, we say that the ${\cal PT}$ symmetry is unbroken.
Conversely, if the solution does not possess ${\cal PT}$ symmetry,
we say that the ${\cal PT}$ symmetry is broken.}. It is easy to
show that if the ${\cal PT}$ symmetry of a Hamiltonian $H$ is
unbroken, then the spectrum of $H$ is real. The proof is short and
goes as follows: Assume that a Hamiltonian $H$ possesses ${\cal
PT}$ symmetry (that is, that $H$ commutes with the ${\cal PT}$
operator), and that if $\phi$ is an eigenstate of $H$ with
eigenvalue $E$, then it is simultaneously an eigenstate of ${\cal
PT}$ with eigenvalue $\lambda$:
\begin{equation}
H\phi=E\phi\quad{\rm and}\quad{\cal PT}\phi=\lambda\phi.
\label{e2}
\end{equation}

We begin by showing that the eigenvalue $\lambda$ is a pure phase.
Multiplying ${\cal PT}\phi=\lambda\phi$ on the left by ${\cal PT}$
and using the fact that ${\cal P}$ and ${\cal T}$ commute and that
${\cal P}^2={\cal T}^2=1$ we conclude that
$\phi=\lambda^*\lambda\phi$ and thus $\lambda=e^{i\alpha}$ for
some real $\alpha$. Next, we introduce the convention that is used
throughout this paper. Without loss of generality we replace the
eigenstate $\phi$ by $e^{-i\alpha/2} \phi$ so that its eigenvalue
under the operator ${\cal PT}$ is unity:
\begin{equation}
\quad {\cal PT}\phi=\phi. \label{e3}
\end{equation}
Let us turn to the eigenvalue equation $H\phi=E\phi$. We multiply
this equation on the left by ${\cal PT}$ and use the fact that
$[{\cal PT},H]=0$ to obtain $E\phi=E^*\phi$. Hence, $E=E^*$ and
the eigenvalue $E$ is real.

The crucial assumption in this argument is that $\phi$ is
simultaneously an eigenstate of $H$ and ${\cal PT}$. In quantum
mechanics if a linear operator $X$ commutes with the Hamiltonian
$H$, then the eigenstates of $H$ are also eigenstates of $X$.
However, we emphasize that the operator ${\cal PT}$ is not linear
(it is antilinear) and thus we must make the extra assumption that
the ${\cal PT}$ symmetry of $H$ is unbroken; that is, that $\phi$
is simultaneously an eigenstate of $H$ and ${\cal PT}$. This extra
assumption is nontrivial because it is not easy to determine {\it
a priori} whether the ${\cal PT}$ symmetry of a particular
Hamiltonian $H$ is broken or unbroken. For the Hamiltonian $H$ in
(\ref{e1}) the ${\cal PT}$ symmetry is unbroken when
$\epsilon\geq0$ and it is broken when $\epsilon<0$. Note that the
conventional Hermitian Hamiltonian for the quantum mechanical
harmonic oscillator lies at the boundary of the unbroken and the
broken regimes. Recently, Dorey {\em et al.} proved rigorously
that the spectrum of $H$ in (\ref{e1}) is real and positive
\cite{r3} in the region $\epsilon\geq0$. Many other ${\cal
PT}$-symmetric Hamiltonians for which space-time reflection
symmetry is not broken have been investigated, and the spectra of
these Hamiltonians have also been shown to be real and positive
\cite{r4}.

While it is useful to show that a given non-Hermitian ${\cal
PT}$-symmetric Hamiltonian operator has a positive real spectrum,
the urgent question that must be answered is whether such a
Hamiltonian defines a physical theory of quantum mechanics. By a
{\it physical theory} we mean that there is a Hilbert space of
state vectors and that this Hilbert space has an inner product
with a positive norm. In the theory of quantum mechanics we
interpret the norm of a state as a probability and this
probability must be positive. Furthermore, we must show that the
time evolution of the theory is unitary. This means that as a
state vector evolves in time the probability does not leak away.

It is not at all obvious whether a Hamiltonian such as $H$ in
(\ref{e1}) gives rise to a consistent quantum theory. Indeed,
while past investigations of this Hamiltonian have shown that the
spectrum is entirely real and positive when $\epsilon\geq0$, it
appeared that one inevitably encountered the severe problem of
dealing with Hilbert spaces endowed with indefinite metrics
\cite{r5}. In this paper we will identify a new symmetry that all
${\cal PT}$-symmetric Hamiltonians having an unbroken ${\cal
PT}$-symmetry possess. We denote the operator representing this
symmetry by ${\cal C}$ because the properties of this operator
resemble those of the charge conjugation operator in particle
physics. This will allow us to introduce an inner product
structure associated with ${\cal CPT}$ conjugation for which the
norms of quantum states are positive definite. We will see that
${\cal CPT}$ symmetry is an alternative to the conventional
Hermiticity requirement; it introduces the new concept of a {\sl
dynamically determined} inner product (one that is defined by the
Hamiltonian itself). As a consequence, we will extend the
Hamiltonian and its eigenstates into the complex domain so that
the associated eigenvalues are real and the underlying dynamics is
unitary.

\section{Construction of the ${\cal C}$ Operator}
\label{s2} We begin by summarizing the mathematical properties of
the solution to the Sturm-Liouville differential equation
eigenvalue problem
\begin{equation}
-\phi_n''(x)+x^2(ix)^\epsilon\phi_n(x)=E_n\phi_n(x) \label{e4}
\end{equation}
associated with the Hamiltonian $H$ in (\ref{e1}). The
differential equation (\ref{e4}) must be imposed on an infinite
contour in the complex-$x$ plane. For large $|x|$ this contour
lies in wedges that are placed symmetrically with respect to the
imaginary-$x$ axis \cite{r2}. The boundary conditions on the
eigenfunctions are that $\phi(x)\to0$ exponentially rapidly as
$|x|\to\infty$ on the contour. For $0\leq\epsilon<2$, the contour
may be taken to be the real axis.

When $\epsilon\geq0$, the Hamiltonian has an unbroken ${\cal PT}$
symmetry. Thus, the eigenfunctions $\phi_n(x)$ are simultaneously
eigenstates of the ${\cal PT}$ operator: ${\cal
PT}\phi_n(x)=\lambda_n\phi_n(x)$. As we argued above, $\lambda_n$
is a pure phase and, without loss of generality, for each $n$ this
phase can be absorbed into $\phi_n(x)$ by a multiplicative
rescaling so that the new eigenvalue is unity:
\begin{equation}
{\cal PT}\phi_n(x)=\phi_n^*(-x)=\phi_n(x). \label{e5}
\end{equation}

There is strong evidence that, when properly normalized, the
eigenfunctions $\phi_n(x)$ are complete. The coordinate-space
statement of completeness (for real $x$ and $y$) reads
\begin{equation}
\sum_n(-1)^n\phi_n(x)\phi_n(y)=\delta(x-y). \label{e6}
\end{equation}
This is a nontrivial result that has been verified numerically to
extremely high accuracy (twenty decimal places) \cite{r6,r7}. Note
that there is a factor of $(-1)^n$ in the sum. This unusual factor
does not appear in conventional quantum mechanics. The presence of
this factor is explained in the following discussion of
orthonormality [see (\ref{e8})].

Here is where we encounter the underlying problem associated with
non-Hermitian ${\cal PT}$-symmetric Hamiltonians. There seems to
be a natural choice for the inner product of two functions $f(x)$
and $g(x)$:
\begin{equation}
(f,g)\equiv\int dx\,[{\cal PT}f(x)]g(x), \label{e7}
\end{equation}
where ${\cal PT}f(x)=[f(-x)]^*$ and the integral is taken over the
contour described above in the complex-$x$ plane. The apparent
advantage of this inner product is that the associated norm
$(f,f)$ is independent of the overall phase of $f(x)$ and is
conserved in time. Phase independence is desired because in the
theory of quantum mechanics the objective is to construct a space
of rays to represent quantum mechanical states. With respect to
this inner product the eigenfunctions $\phi_m(x)$ and $\phi_n(x)$
of $H$ in (\ref{e1}) are orthogonal for $n\neq m$. However, when
$m=n$ the norm is evidently {\it not positive}:
\begin{equation}
(\phi_m,\phi_n)=(-1)^n\delta_{mn}. \label{e8}
\end{equation}
This result is apparently true for all values of $\epsilon$ in
(\ref{e4}) and it has been verified numerically to extremely high
precision. Because the norms of the eigenfunctions alternate in
sign, the Hilbert space metric associated with the ${\cal PT}$
inner product $(\cdot,\cdot)$ is indefinite. This split signature
(sign alternation) is a {\it generic} feature of the ${\cal PT}$
inner product. Extensive numerical calculations verify that the
formula in (\ref{e8}) holds for all $\epsilon\geq0$.

Despite the lack of positivity of the inner product, we proceed
with the usual analysis that one would perform for any
Sturm-Liouville problem of the form $H\phi_n=E_n\phi_n$. First, we
use the inner product formula (\ref{e8}) to verify that (\ref{e6})
is the representation of the unity operator. That is, we verify
that
\begin{equation}
\int dy\,\delta(x-y)\delta(y-z)=\delta(x-z). \label{e9}
\end{equation}
Second, we reconstruct the parity operator ${\cal P}$ in terms of
the eigenstates. The parity operator in position space is ${\cal
P}(x,y)=\delta(x+y)$, so from (\ref{e6}) we get
\begin{equation}
{\cal P}(x,y)=\sum_n(-1)^n\phi_n(x)\phi_n(-y). \label{e10}
\end{equation}
By virtue of (\ref{e8}) the square of the parity operator is
unity: ${\cal P}^2 =1$.

Third, we reconstruct the Hamiltonian $H$ in coordinate space:
\begin{equation}
H(x,y)=\sum_n(-1)^nE_n\phi_n(x)\phi_n(y). \label{e11}
\end{equation}
Using (\ref{e6}) - (\ref{e8}) it is easy to see that this
Hamiltonian satisfies $H\phi_n(x)=E_n\phi_n(x)$. Fourth, we
construct the coordinate-space Green's function $G(x,y)$:
\begin{equation}
G(x,y)=\sum_n(-1)^n\frac{1}{E_n}\phi_n(x)\phi_n(y). \label{e12}
\end{equation}
Note that the Green's function is the functional inverse of the
Hamiltonian; that is, $G$ satisfies the equation
\begin{equation}
\int dy\,H(x,y)G(y,z)=\left[-\frac{d^2}{dx^2}
+x^2(ix)^\epsilon\right]G(x,z)= \delta(x-z). \label{e13}
\end{equation}
While the time-independent Schr\"odinger equation (\ref{e4})
cannot be solved analytically, the differential equation for
$G(x,z)$ in (\ref{e13}) {\it can} be solved exactly and in closed
form \cite{r7}. The technique is to consider the case
$0<\epsilon<2$ so that we may treat $x$ as real and then to
decompose the $x$ axis into two regions, $x>z$ and $x<z$. We can
solve the differential equation in each of these regions in terms
of Bessel functions. Then, using this coordinate-space
representation of the Green's function, we construct an exact
closed-form expression for the {\it spectral zeta function} (sum
of the inverses of the energy eigenvalues). To do so we set $y=x$
in $G(x,y)$ and use (\ref{e8}) to integrate over $x$. For all
$\epsilon>0$ we obtain \cite{r7}
\begin{equation}
\sum_n {1\over
E_n}=\left[1+{\cos\left({3\epsilon\pi\over2\epsilon+8}\right)\sin
\left({\pi\over4+\epsilon}\right)\over\cos\left({\epsilon\pi\over4+2\epsilon}
\right)\sin\left({3\pi\over4+\epsilon}\right)}\right]{\Gamma\left({1\over4+
\epsilon}\right)\Gamma\left({2\over4+\epsilon}\right)\Gamma\left({\epsilon\over
4+\epsilon}\right)\over(4+\epsilon)^{4+2\epsilon\over4+\epsilon}\Gamma\left({1+
\epsilon\over4+\epsilon}\right)\Gamma\left({2+\epsilon\over4+\epsilon}\right)}.
\label{e14}
\end{equation}

Having presented these general Sturm-Liouville constructions, we
now address the crucial question of whether a ${\cal
PT}$-symmetric Hamiltonian defines a physically viable quantum
mechanics or whether it merely provides an intriguing
Sturm-Liouville eigenvalue problem. The apparent difficulty with
formulating a quantum theory is that the vector space of quantum
states is spanned by energy eigenstates, of which half have norm
$+1$ and half have norm $-1$. Because the norm of the states
carries a probabilistic interpretation in standard quantum theory,
the existence of an indefinite metric in (\ref{e8}) seems to be a
serious obstacle.

The situation here in which half of the energy eigenstates have
positive norm and half have negative norm is analogous to the
problem that Dirac encountered in formulating the spinor wave
equation in relativistic quantum theory \cite{r8}. Following
Dirac's approach, we attack the problem of an indefinite norm by
finding a physical interpretation for the negative norm states. We
claim that in {\it any} theory having an unbroken ${\cal PT}$
symmetry there exists a symmetry of the Hamiltonian connected with
the fact that there are equal numbers of positive-norm and
negative-norm states. To describe this symmetry we construct a
linear operator denoted by ${\cal C}$ and represented in position
space as a sum over the energy eigenstates of the Hamiltonian
\cite{r9}:
\begin{equation}
{\cal C}(x,y)=\sum_n\phi_n(x)\phi_n(y). \label{e15}
\end{equation}

As stated earlier, the properties of this new operator ${\cal C}$
are nearly identical to those of the charge conjugation operator
in quantum field theory. For example, we can use equations
(\ref{e6}) - (\ref{e8}) to verify that the square of ${\cal C}$ is
unity (${\cal C}^2=1$):
\begin{equation}
\int dy\,{\cal C}(x,y){\cal C}(y,z)=\delta(x-z). \label{e16}
\end{equation}
Thus, the eigenvalues of ${\cal C}$ are $\pm1$. Also, ${\cal C}$
commutes with the Hamiltonian $H$. Therefore, since ${\cal C}$ is
linear, the eigenstates of $H$ have definite values of ${\cal C}$.
Specifically, if the energy eigenstates satisfy (\ref{e8}), then
we have ${\cal C}\phi_n=(-1)^n\phi_n$ because
$${\cal C}\phi_n(x)=\int dy\,{\cal C}(x,y)\phi_n(y)=\sum_m\phi_m(x)\int dy\,
\phi_m(y)\phi_n(y).$$ We then use $\int
dy\,\phi_m(y)\phi_n(y)=(\phi_m,\phi_n)$ according to our
convention. We conclude that ${\cal C}$ is the operator observable
that represents the measurement of the signature of the ${\cal
PT}$ norm of a state \footnote{The ${\cal PT}$ norm of a state
determines its parity type \cite{BBHM}. We can regard ${\cal C}$
as representing the operator that determines the ${\cal C}$ charge
of the state. Quantum states having opposite ${\cal C}$ charge
possess opposite parity type.}.

Note that the operators ${\cal P}$ and ${\cal C}$ are distinct
square roots of the unity operator $\delta(x-y)$. That is, while
${\cal P}^2=1$ and ${\cal C}^2= 1$, ${\cal P}$ and ${\cal C}$ are
not identical. Indeed, the parity operator ${\cal P}$ is real,
while ${\cal C}$ is complex \footnote{The parity operator in
coordinate space is explicitly real ${\cal P}(x,y)=\delta(x+y)$;
the operator ${\cal C}(x,y)$  is complex because it is a sum of
products of complex functions, as we see in (\ref{e15}). The
complexity of the ${\cal C}$ operator can be seen explicitly in
perturbative calculations of ${\cal C}(x,y)$ \cite{AAA}.}.
Furthermore, these two operators do not commute; in the position
representation
\begin{equation}
({\cal CP})(x,y)=\sum_n\phi_n(x)\phi_n(-y)\quad{\rm but}\quad
({\cal PC})(x,y)= \sum_n\phi_n(-x)\phi_n(y), \label{e17}
\end{equation}
which shows that ${\cal CP}=({\cal PC})^*$. However, ${\cal C}$
{\it does} commute with ${\cal PT}$.

Finally, having obtained the operator ${\cal C}$ we define a new
inner product structure having {\it positive definite} signature
by
\begin{equation}
\langle f|g\rangle\equiv\int_{\rm{C}} dx\,[{\cal CPT}f(x)]g(x).
\label{e18}
\end{equation}
Like the ${\cal PT}$ inner product (\ref{e7}), this inner product
is phase independent and conserved in time. This is because the
time evolution operator, just as in ordinary quantum mechanics, is
$e^{iHt}$. The fact that $H$ commutes with the ${\cal PT}$ and the
${\cal CPT}$ operators implies that both inner products,
(\ref{e7}) and (\ref{e18}), remain time independent as the states
evolve in time. However, unlike (\ref{e7}), the inner product
(\ref{e18}) is positive definite because ${\cal C}$ contributes
$-1$ when it acts on states with negative ${\cal PT}$ norm. In
terms of the ${\cal CPT}$ conjugate, the completeness condition
(\ref{e4}) reads
\begin{equation}
\sum_n \phi_n(x)[{\cal CPT} \phi_n(y)]=\delta(x-y). \label{e19}
\end{equation}
Unlike the inner product of conventional quantum mechanics, the
${\cal CPT}$ inner product (\ref{e19}) is {\it dynamically
determined}; it depends implicitly on the choice of Hamiltonian.

The operator ${\cal C}$ does not exist as a distinct entity in
conventional quantum mechanics. Indeed, if we allow the parameter
$\epsilon$ in (\ref{e1}) to tend to zero, the operator ${\cal C}$
in this limit becomes identical to ${\cal P}$. Thus, in this limit
the ${\cal CPT}$ operator becomes ${\cal T}$, which is just
complex conjugation. As a consequence, the inner product
(\ref{e18}) defined with respect to the ${\cal CPT}$ conjugation
reduces to the complex conjugate inner product of conventional
quantum mechanics when $\epsilon\to0$. Similarly, in this limit
(\ref{e19}) reduces to the usual statement of completeness $\sum_n
\phi_n(x)\phi_n^*(y)=\delta(x-y)$.

Note that the ${\cal CPT}$ inner-product (\ref{e18}) is
independent of the choice of integration contour ${\rm C}$ so long
as ${\rm C}$ lies inside the asymptotic wedges associated with the
boundary conditions for the Sturm-Liouville problem (\ref{e2}).
Path independence is a consequence of Cauchy's theorem and the
analyticity of the integrand. In conventional quantum mechanics,
where the positive-definite inner product has the form $\int dx\,
f^*(x)g(x)$, the integral must be taken along the real axis and
the path of the integration cannot be deformed into the complex
plane because the integrand is not analytic \footnote{Note that if
a function satisfies a linear ordinary differential equation, then
the function is analytic wherever the coefficient functions of the
differential equation are analytic. The Schr\"odinger equation
(\ref{e4}) is linear and its coefficients are analytic except for
a branch cut at the origin; this branch cut can be taken to run up
the imaginary axis. We choose the integration contour for the
inner product (\ref{e8}) so that it does not cross the positive
imaginary axis. Path independence occurs because the integrand of
the inner product (\ref{e8}) is a product of analytic functions.}.
The ${\cal PT}$ inner product (\ref{e7}) shares with (\ref{e18})
the advantage of analyticity and path independence, but suffers
from nonpositivity. We find it surprising that a positive-definite
metric can be constructed using ${\cal CPT}$ conjugation without
disturbing the path independence of the inner-product integral.

Finally, we explain why ${\cal PT}$-symmetric theories are
unitary. Time evolution is determined by the operator $e^{-iHt}$,
whether the theory is expressed in terms of a ${\cal
PT}$-symmetric Hamiltonian or just an ordinary Hermitian
Hamiltonian. To establish the global unitarity of a theory we must
show that as a state vector evolves its norm does not change in
time. If $\psi_0(x)$ is a prescribed initial wave function
belonging to the Hilbert space spanned by the energy eigenstates,
then it evolves into the state $\psi_t(x)$ at time $t$ according
to
$$\psi_t(x)=e^{-iHt}\psi_0(x).$$
With respect to the ${\cal CPT}$ inner product defined in
(\ref{e18}), the norm of the vector $\psi_t(x)$ does not change in
time,
$$\langle\psi_t|\psi_t\rangle=\langle\psi_0|\psi_0\rangle,$$
because the Hamiltonian $H$ commutes with the ${\cal CPT}$
operator. Establishing unitarity at a local level is more
difficult. Here, we must show that in coordinate space, there
exists a local probability density that satisfies a continuity
equation so that the probability does not leak away. This is a
subtle result because the probability current flows about in the
complex plane rather than along the real axis as in conventional
Hermitian quantum mechanics. Preliminary numerical studies indeed
indicate that the continuity equation is fulfilled \cite{CCC}.

\section{Illustrative Example: A $2\times2$ Matrix Hamiltonian}
\label{s3}

We will now illustrate the above results concerning ${\cal
PT}$-symmetric quantum mechanics in a very simple context. To do
so we will consider systems characterized by finite-dimensional
matrix Hamiltonians. In finite-dimensional systems the ${\cal P}$,
${\cal T}$, and ${\cal C}$ operators appear, but there is no
analogue of the boundary conditions associated with
coordinate-space Schr\"odinger equations.

Let us consider the $2\times2$ matrix Hamiltonian
\begin{equation}
H=\left(\begin{array}{cc} re^{i\theta} & s \cr s & re^{-i\theta}
\end{array}\right),
\label{e21}
\end{equation}
where the three parameters $r$, $s$, and $\theta$ are real. This
Hamiltonian is not Hermitian in the usual sense, but it is ${\cal
PT}$ symmetric, where the parity operator is given by \cite{r10}
\begin{equation}
\mathcal{P}=\left(\begin{array}{cc} 0 & 1 \cr 1 & 0
\end{array}\right) \label{e20}
\end{equation}
and ${\cal T}$ performs complex conjugation.

There are two parametric regions for this Hamiltonian. When
$s^2<r^2\sin^2\theta$, the energy eigenvalues form a complex
conjugate pair. This is the region of broken ${\cal PT}$ symmetry.
On the other hand, if $s^2\geq r^2\sin^2\theta$, then the
eigenvalues $\varepsilon_\pm=r\cos\theta\pm\sqrt{s^2-r^2\sin^2
\theta}$ are real. This is the region of unbroken ${\cal PT}$
symmetry. In the unbroken region the simultaneous eigenstates of
the operators $H$ and ${\cal PT}$ are given by
\begin{equation}
|\varepsilon_+\rangle=\frac{1}{\sqrt{2\cos\alpha}}\left(\begin{array}{c}
e^{i\alpha/2}\cr e^{-i\alpha/2}\end{array}\right)\quad{\rm
and}\quad
|\varepsilon_-\rangle=\frac{i}{\sqrt{2\cos\alpha}}\left(\begin{array}{c}
e^{-i\alpha/2}\cr-e^{i\alpha/2}\end{array}\right), \label{e22}
\end{equation}
where we set $\sin\alpha =(r/s)\,\sin\theta$. It is easily
verified that $(\varepsilon_{\pm},\varepsilon_{\pm})=\pm1$ and
that $(\varepsilon_{\pm}, \varepsilon_{\mp})=0$, recalling that
$(u,v)=({\cal PT}u)\cdot v$. Therefore, with respect to the ${\cal
PT}$ inner product, the resulting vector space spanned by energy
eigenstates has a metric of signature $(+,-)$. The condition
$s^2>r^2\sin^2\theta$ ensures that ${\cal PT}$ symmetry is not
broken. If this condition is violated, the states (\ref{e22}) are
no longer eigenstates of ${\cal PT}$ because $\alpha$ becomes
imaginary\footnote{When ${\cal PT}$ symmetry is broken, we find
that the ${\cal PT}$ norm of the energy eigenstate vanishes.}.

Next, we construct the operator ${\cal C}$:
\begin{equation}
{\cal C}=\frac{1}{\cos\alpha}\left(\begin{array}{cc} i\sin\alpha &
1 \cr 1 & -i\sin\alpha \end{array}\right). \label{e23}
\end{equation}
Note that ${\cal C}$ is distinct from $H$ and ${\cal P}$ and has
the key property that
\begin{equation}
{\cal C}|\varepsilon_{\pm}\rangle=\pm |\varepsilon_{\pm}\rangle.
\label{e24}
\end{equation}
The operator ${\cal C}$ commutes with $H$ and satisfies ${\cal
C}^2=1$. The eigenvalues of ${\cal C}$ are precisely the signs of
the ${\cal PT}$ norms of the corresponding eigenstates.

Using the operator ${\cal C}$ we construct the new inner product
structure
\begin{equation}
\langle u|v\rangle=({\cal CPT}u)\cdot v. \label{e25}
\end{equation}
This inner product is positive definite because
$\langle\varepsilon_{\pm}| \varepsilon_{\pm}\rangle=1$. Thus, the
two-dimensional Hilbert space spanned by
$|\varepsilon_\pm\rangle$, with inner product
$\langle\cdot|\cdot\rangle$, has a Hermitian structure with
signature $(+,+)$.

Let us demonstrate explicitly that the ${\cal CPT}$ norm of any
vector is positive. We choose the arbitrary vector
$\psi=\left({a\atop b}\right)$, where $a$ and $b$ are any complex
numbers. We then see that ${\cal T}\psi=\left(a^*\atop
b^*\right)$, that ${\cal PT}\psi=\left(b^*\atop a^*\right)$, and
that ${\cal CPT}\psi={1\over\cos
\alpha}\,\left(a^*+ib^*\sin\alpha\atop b^*-ia^*\sin\alpha\right)$.
Thus, $\langle\psi|\psi\rangle=({\cal
CPT}\psi)\cdot\psi={1\over\cos\alpha}[a^*a+b^*b
+i(b^*b-a^*a)\sin\alpha]$. Now let $a=x+iy$ and $b=u+iv$, where
$x$, $y$, $u$, and $v$ are real. Then
\begin{equation}
\langle\psi|\psi\rangle={1\over\cos\alpha}\left(x^2+v^2+2xv\sin\alpha
+y^2+u^2-2yu\sin\alpha\right), \label{e26}
\end{equation}
which is explicitly positive and vanishes only if $x=y=u=v=0$.

Recalling that $\langle u|$ denotes the ${\cal CPT}$-conjugate of
$|u\rangle$, the completeness condition reads
\begin{equation}
|\varepsilon_+\rangle\langle\varepsilon_+|+|\varepsilon_-\rangle\langle
\varepsilon_-|=\left(\begin{array}{cc} 1 & 0 \cr 0 &
1\end{array}\right). \label{e27}
\end{equation}
Furthermore, using the ${\cal CPT}$ conjugate
$\langle\varepsilon_\pm|$, we can express ${\cal C}$ in the form
${\cal C}=|\varepsilon_+\rangle\langle \varepsilon_+| -
|\varepsilon_-\rangle\langle\varepsilon_-|,$ as opposed to the
representation in (\ref{e15}), which uses the ${\cal PT}$
conjugate.

In general, an observable in this theory is represented by a
${\cal CPT}$ invariant operator; that is, one that commutes with
${\cal CPT}$. Thus, if ${\cal CPT}$ symmetry is not broken, the
eigenvalues of the observable are real. The operator ${\cal C}$
satisfies this requirement, and hence it is an observable. For the
two-state system, if we set $\theta=0$, then the Hamiltonian
(\ref{e21}) becomes Hermitian. However, the operator ${\cal C}$
then reduces to the parity operator ${\cal P}$. As a consequence,
the requirement of ${\cal CPT}$ invariance reduces to the standard
condition of Hermiticity for a symmetric matrix, namely, that
$H=H^*$. This is why the hidden symmetry ${\cal C}$ was not
noticed previously. The operator ${\cal C}$ emerges only when we
extend a real symmetric Hamiltonian into the complex domain.

We have also calculated the ${\cal C}$ operator in
infinite-dimensional quantum mechanical models. For an $x^2+ix^3$
potential ${\cal C}$ can be obtained from the summation in
(\ref{e15}) using perturbative methods and for an $x^2-x^4$
potential ${\cal C}$ can be calculated using nonperturbative
methods \cite{AAA}.

\section{Applications and Possible Observable Consequences}
\label{s4} In summary, we have described an alternative to the
axiom of Hermiticity in quantum mechanics; we call this new
requirement ${\cal PT}$ invariance. In quantum field theory,
Hermiticity, Lorentz invariance, and a positive spectrum are
crucial for establishing ${\cal CPT}$ invariance \cite{r11}. Here,
we have established the converse of the ${\cal CPT}$ theorem in
the following limited sense: We assume that the Hamiltonian
possesses space-time reflection symmetry, and that this symmetry
is not broken. From these assumptions, we know that the spectrum
is real and positive and we construct an operator ${\cal C}$ that
is like the charge conjugation operator. We show that quantum
states in this theory have positive norms with respect to ${\cal
CPT}$ conjugation. In effect, we replace the mathematical
condition of Hermiticity, whose physical content is somewhat
remote and obscure, by the physical condition of space-time and
charge-conjugation symmetry. These symmetries ensure the reality
of the spectrum of the Hamiltonian in complex quantum theories.

Could non-Hermitian, ${\cal PT}$-symmetric Hamiltonians be used to
describe experimentally observable phenomena? Non-Hermitian
Hamiltonians have {\it already} been used to describe interacting
systems. For example in 1959, Wu showed that the ground state of a
Bose system of hard spheres is described by a non-Hermitian
Hamiltonian \cite{r12}. Wu found that the ground-state energy of
this system is real and conjectured that all of the energy levels
were real. In 1992, Hollowood showed that even though the
Hamiltonian of a complex Toda lattice is non-Hermitian, the energy
levels are real \cite{r13}. Non-Hermitian Hamiltonians of the form
$H=p^2+ix^3$ also arise in various Reggeon field theory models
that exhibit real positive spectra \cite{r14}. In each of these
cases the fact that a non-Hermitian Hamiltonian had a real
spectrum appeared mysterious at the time, but now the explanation
is simple: In each of these cases it is easy to show that the
non-Hermitian Hamiltonian is ${\cal PT}$-symmetric. That is, the
Hamiltonian in each case is constructed so that the position
operator $x$ or the field operator $\phi$ is always multiplied by
$i$.

An experimental signal of a complex Hamiltonian might be found in
the context of condensed matter physics. Consider the complex
crystal lattice whose potential is given by $V(x)=i\sin\,x$. While
the Hamiltonian $H=p^2+i\sin\,x$ is not Hermitian, it is ${\cal
PT}$-symmetric, and all of the energy bands are {\it real}.
However, at the edge of the bands the wave function of a particle
in such a lattice is always bosonic ($2\pi$-periodic) and, unlike
the case of ordinary crystal lattices, the wave function is never
fermionic ($4\pi$-periodic) \cite{r15}. Direct observation of such
a band structure would give unambiguous evidence of a ${\cal
PT}$-symmetric Hamiltonian.

There are many opportunities for the use of non-Hermitian
Hamiltonians in the study of quantum field theory. For example, a
scalar quantum field theory with a cubic self-interaction
described by the Lagrangian ${\cal L}={1\over2}(\nabla
\varphi)^2+{1\over2}m^2\varphi^2+g\varphi^3$ is physically
unacceptable because the energy spectrum is not bounded below.
However, the cubic scalar quantum field theory that corresponds to
$H$ in (\ref{e1}) with $\epsilon=1$ is given by the Lagrangian
density ${\cal L}={1\over2}(\nabla\varphi)^2+{1\over2}m^2\varphi
^2+ig\varphi^3$. This is a new, physically acceptable quantum
field theory. Moreover, the theory that corresponds to $H$ in
(\ref{e1}) with $\epsilon=2$ is described by the Lagrangian
density
\begin{equation}
{\cal
L}={1\over2}(\nabla\varphi)^2+{1\over2}m^2\varphi^2-{1\over4}g\varphi^4.
\label{e28}
\end{equation}
What is remarkable about this ``wrong-sign'' field theory is that,
in addition to the energy spectrum being real and positive, the
one-point Green's function (the vacuum expectation value of the
field $\varphi$) is {\it nonzero} \cite{r16}. Furthermore, the
field theory is renormalizable, and in four dimensions is
asymptotically free (and thus nontrivial) \cite{r17}. Based on
these features of the theory, we believe that the theory may
provide a useful setting to describe the dynamics of the Higgs
sector in the standard model.

Other field theory models whose Hamiltonians are non-Hermitian and
${\cal PT}$-symmetric have also been studied. For example, ${\cal
PT}$-symmetric electrodynamics is particularly interesting because
it is asymptotically free (unlike ordinary electrodynamics) and
because the direction of the Casimir force is the negative of that
in ordinary electrodynamics \cite{r18}. This theory is remarkable
because it can determine its own coupling constant. Supersymmetric
${\cal PT}$-symmetric quantum field theories have also been
studied \cite{r19}.

We have found that ${\cal PT}$-symmetric quantum theories exhibit
surprising and new phenomena. For example, when $g$ is
sufficiently small, the $-g\varphi ^4$ theory described by the
Lagrangian (\ref{e28}) possesses bound states (the conventional
$g\varphi^4$ theory does not because the potential is repulsive).
The bound states occur for all dimensions $0\leq D<3$ \cite{r20},
but for purposes of illustration we describe the bound states in
the context of one-dimensional quantum field theory (quantum
mechanics). For the conventional quantum mechanical anharmonic
oscillator, which is described by the Hamiltonian
\begin{equation}
H={1\over2}p^2+{1\over2}m^2x^2+{1\over4}gx^4\qquad(g>0),
\label{e29}
\end{equation}
the small-$g$ Rayleigh-Schr\"odinger perturbation series for the
$k$th energy level $E_k$ is
\begin{eqnarray}
E_k\sim m\left[k+{1\over2}+{3\over4}(2k^2+2k+1)\nu+{\rm
O}(\nu^2)\right] \qquad(\nu\to0^+), \label{e30}
\end{eqnarray}
where $\nu=g/(4m^3)$. The {\it renormalized mass} $M$ is defined
as the first excitation above the ground state: $M\equiv
E_1-E_0\sim m[1+3\nu+{\rm O}(\nu^2 )]$ as $\nu\to0^+$.

To determine if the two-particle state is bound, we examine the
second excitation above the ground state using (\ref{e30}). We
define
\begin{eqnarray}
B_2\equiv E_2-E_0\sim m\left[2+9\nu+{\rm
O}(\nu^2)\right]\qquad(\nu\to0^+). \label{e31}
\end{eqnarray}
If $B_2<2M$, then a two-particle bound state exists and the
(negative) binding energy is $B_2-2M$. If $B_2>2M$, then the
second excitation above the vacuum is interpreted as an unbound
two-particle state. We see from (\ref{e31}) that in the
small-coupling region, where perturbation theory is valid, the
conventional anharmonic oscillator does not possess a bound state.
Indeed, using WKB, variational methods, or numerical calculations,
one can show that there is no two-particle bound state for any
value of $g>0$. Because there is no bound state the $gx^4$
interaction may be considered to represent a repulsive force
\footnote{In general, a repulsive force in a quantum field theory
is represented by an energy dependence in which the energy of a
two-particle state decreases with separation. The conventional
anharmonic oscillator Hamiltonian corresponds to a field theory in
one space-time dimension, where there cannot be any spatial
dependence. In this case the repulsive nature of the force is
understood to mean that the energy $B_2$ needed to create two
particles at a given time is more than twice the energy $M$ needed
to create one particle.}.

We obtain the perturbation series for the non-Hermitian, ${\cal
PT}$-symmetric Hamiltonian
\begin{equation}
H={1\over2}p^2+{1\over2}m^2x^2-{1\over4}gx^4\qquad(g>0),
\label{e32}
\end{equation}
from the perturbation series for the conventional anharmonic
oscillator by replacing $\nu$ with $-\nu$. Thus, while the
conventional anharmonic oscillator does not possess a two-particle
bound state, the ${\cal PT}$-symmetric oscillator does indeed
possess such a state. We measure the binding energy of this state
in units of the renormalized mass $M$ and we define the {\it
dimensionless} binding energy $\Delta_2$ by
\begin{eqnarray}
\Delta_2\equiv{B_2-2M\over M}\sim-3\nu+{\rm
O}(\nu^2)\qquad(\nu\to0^+). \label{e33}
\end{eqnarray}
This bound state disappears when $\nu$ increases beyond
$\nu=0.0465\ldots$. As $\nu$ continues to increase, $\Delta_2$
reaches a maximum value of $0.427$ at $\nu=0.13$ and then
approaches the limiting value $0.28$ as $\nu\to\infty$.

In the ${\cal PT}$-symmetric anharmonic oscillator, there are not
only two-particle bound states for small coupling constant but
also $k$-particle bound states for all $k\geq2$. The dimensionless
binding energies are
\begin{eqnarray}
\Delta_k\equiv(B_k-kM)/M\sim-3k(k-1)\nu/2+{\rm
O}(\nu^2)\qquad(\nu\to0+). \label{e34}
\end{eqnarray}
The key feature of this equation is that the coefficient of $\nu$
is negative. Since the dimensionless binding energy becomes
negative as $\nu$ increases from $0$, there is a $k$-particle
bound state. The higher multiparticle bound states cease to be
bound for smaller values of $\nu$; starting with the
three-particle bound state, the binding energy of these states
becomes positive as $\nu$ increases past $0.039$, $0.034$,
$0.030$, and $0.027$.

Thus, for any value of $\nu$ there are always a finite number of
bound states and an infinite number of unbound states. The number
of bound states decreases with increasing $\nu$ until there are no
bound states at all. There is a range of $\nu$ for which there are
only two- and three-particle bound states. This situation is
analogous to the physical world in which one observes only states
of two and three bound quarks. In this range of $\nu$ if one has
an initial state containing a number of particles (renormalized
masses), these particles will clump together into bound states,
releasing energy in the process. Depending on the value of $\nu$,
the final state will consist either of two- or of three-particle
bound states, whichever is energetically favored. Note also that
there is a special value of $\nu$ for which two- and
three-particle bound states can exist in thermodynamic
equilibrium.

How does a $g\varphi^3$ theory compare with a $g\varphi^4$ theory?
A $g\varphi ^3$ theory has an attractive force. The bound states
that arise as a consequence of this force can be found by using
the Bethe-Salpeter equation. However, the $g \varphi^3$ field
theory is unacceptable because the spectrum is not bounded below.
If we replace $g$ by $ig$, the spectrum becomes real and positive,
but now the force becomes repulsive and there are no bound states.
The same is true for a two-scalar theory with interaction of the
form $ig\varphi^2\chi$. This latter theory is an acceptable model
of scalar electrodynamics, but has no analog of positronium.

Another feature of ${\cal PT}$-symmetric quantum field theory that
distinguishes it from the conventional quantum field theory lies
in the commutation relation between the ${\cal P}$ and ${\cal C}$
operators. Specifically, if we write ${\cal C}={\cal C}_{\rm
R}+i{\cal C}_{\rm I}$, where ${\cal C}_{\rm R}$ and ${\cal C}_{\rm
I}$ are real, then ${\cal C}_{\rm R}{\cal P}={\cal P}{\cal C}_{\rm
R}$ and ${\cal C}_{\rm I}{\cal P}=-{\cal P}{\cal C}_{\rm I}$.
These commutation and anticommutation relations suggest the
possibility of interpreting ${\cal PT}$-symmetric quantum field
theory as describing both bosonic and fermionic degrees of
freedom, an idea analogous to the supersymmetric quantum theories.
The distinction here, however, is that the supersymmetry can be
broken; that is, bosonic and fermionic counterparts can have
different masses without breaking the ${\cal PT}$ symmetry.
Therefore, another possible observable experimental consequence
might be the breaking of the supersymmetry.

\section{Concluding Remarks}
\label{s5} We have argued in this paper that there is an
alternative to the axiom of standard quantum mechanics that the
Hamiltonian must be Hermitian. We have shown that the axiom of
Hermiticity may be replaced by the more physical condition of
${\cal PT}$ (space-time reflection) symmetry. Space-time
reflection symmetry is distinct from the condition of Hermiticity,
so it is possible to consider new kinds of quantum theories, such
as quantum field theories whose self-interaction potentials are
$ig\varphi^3$ or $-g\varphi^4$. Such theories have previously been
thought to be mathematically and physically unacceptable because
the spectrum might not be real and because the time evolution
might not be unitary.

These new kinds of theories can be thought of as extensions of
ordinary quantum mechanics into the complex plane; that is,
continuations of real symmetric Hamiltonians to complex
Hamiltonians. The idea of analytically continuing a Hamiltonian
was first discussed in 1952 by Dyson, who argued heuristically
that perturbation theory for quantum electrodynamics is divergent
\cite{DY}. Dyson's argument involves rotating the electric charge
$e$ into the complex plane $e\to ie$. Applied to the quantum
anharmonic oscillator, whose Hamiltonian is given in (\ref{e29}),
Dyson's argument would go as follows: If the coupling constant $g$
is continued in the complex-$g$ plane to $-g$, then the potential
is no longer bounded below, so the resulting theory has no ground
state. Thus, the ground-state energy $E_0(g)$ has an abrupt
transition at $g=0$. If we represent $E_0(g)$ as a series in
powers of $g$, this series must have a zero radius of convergence
because $E_0(g)$ has a singularity at $g=0$ in the
complex-coupling-constant plane. Hence, the perturbation series
must diverge for all $g\neq0$. While the perturbation series does
indeed diverge, this heuristic argument is flawed because the
spectrum of the Hamiltonian (\ref{e32}) that is obtained remains
ambiguous until the boundary conditions that the wave functions
must satisfy are specified. The spectrum depends crucially on how
this Hamiltonian with a negative coupling constant is obtained.

There are two ways to obtain the Hamiltonian (\ref{e32}). First,
one can substitute $g=|g|e^{i\theta}$ into the Hamiltonian
(\ref{e29}) and rotate from $\theta=0$ to $\theta=\pi$. Under this
rotation, the ground-state energy $E_0(g)$ becomes complex.
Evidently, $E_0(g)$ is real and positive when $g>0$ and complex
when $g<0$ \footnote{Rotating from $\theta=0$ to $\theta =-\pi$,
we obtain the same Hamiltonian as in (\ref{e32}) but the spectrum
is the complex conjugate of the spectrum obtained when we rotate
from $\theta=0$ to $\theta= \pi$.}. Second, one can obtain
(\ref{e32}) as a limit of the Hamiltonian
\begin{equation}
H={1\over2}p^2+{1\over2}m^2x^2+{1\over4}gx^2(ix)^\epsilon\qquad(g>0)
\label{e35}
\end{equation}
as $\epsilon:0\to2$. The spectrum of this Hamiltonian is real,
positive, and discrete. The spectrum of the limiting Hamiltonian
(\ref{e32}) obtained in this manner is similar in structure to
that of the Hamiltonian in (\ref{e29}).

How can the Hamiltonian (\ref{e32}) possess two such astonishingly
different spectra? The answer lies in the boundary conditions
satisfied by the wave functions $\phi_n(x)$. In the first case, in
which $\theta={\rm arg}\,g$ is rotated in the complex-$g$ plane
from $0$ to $\pi$, $\psi_n(x)$ vanishes in the complex-$x$ plane
as $|x|\to\infty$ inside the wedges $-\pi/3<{\rm arg}\,x<0$ and
$-4\pi/3<{\rm arg}\,x<-\pi$. In the second case, in which the
exponent $\epsilon$ ranges from $0$ to $2$, $\phi_n(x)$ vanishes
in the complex-$x$ plane as $|x|\to\infty$ inside the wedges
$-\pi/3<{\rm arg}\,x<0$ and $-\pi<{\rm arg} \,x<-2\pi/3$. In this
second case the boundary conditions hold in wedges that are
symmetric with respect to the imaginary axis; these boundary
conditions enforce the $\cal PT$ symmetry of $H$ and are
responsible for the reality of the energy spectrum.

Apart from the spectra, there is another striking difference
between the two theories corresponding to $H$ in (\ref{e32}). The
one-point Green's function $G_1(g)$ is defined as the expectation
value of the operator $x$ in the ground-state wave function
$\phi_0(x)$,
\begin{equation}
G_1(g)=\langle0|x|0\rangle/\langle0|0\rangle\equiv\int_C dx\,x
\psi_0^2(x)\Bigm/\int_C dx\,\psi_0^2(x), \label{e36}
\end{equation}
where $C$ is a contour that lies in the asymptotic wedges
described above. The value of $G_1(g)$ for $H$ in (\ref{e32})
depends on the limiting process by which we obtain $H$. If we
substitute $g=g_0e^{i\theta}$ into the Hamiltonian (\ref{e29}) and
rotate from $\theta=0$ to $\theta=\pi$, we find by an elementary
symmetry argument that $G_1(g)=0$ for all $g$ on the semicircle in
the complex-$g$ plane. Thus, this rotation in the complex-$g$
plane preserves parity symmetry ($x\to-x$). However, if we define
$H$ in (\ref{e32}) by using the Hamiltonian in (\ref{e35}) and by
allowing $\epsilon$ to range from $0$ to $2$, we find that
$G_1(g)\neq0$. Indeed, $G_1(g)\neq0$ for {\it all} values of
$\epsilon>0$. Thus, in this theory ${\cal PT}$ symmetry
(reflection about the imaginary axis, $x\to-x^*$) is preserved,
but parity symmetry is permanently broken. We believe that this
means that one might be able to describe the dynamics of the Higgs
sector by using a $-g\varphi^4$ quantum field theory.

\section*{Acknowledgement}
DCB gratefully acknowledges financial support from The Royal
Society. This work was supported in part by the U.S.~Department of
Energy.

\end{document}